\documentclass[prd,english,aps,showpacs,nofootinbib,preprint,eqsecnum]{revtex4-1}
\usepackage{graphicx,color,amsmath,amsxtra}
\usepackage{epsf}
\usepackage{amssymb}
\usepackage{enumerate}
\usepackage{hhline}
\usepackage{array}
\usepackage{tabularx}

\begin{document}

\title{Singularity phenomena in viable $f(R)$ gravity}

\author{Chung-Chi Lee$^{1}$\footnote{E-mail address:
g9522545@oz.nthu.edu.tw},
Chao-Qiang Geng$^{1,2}$\footnote{E-mail address: geng@phys.nthu.edu.tw},
and Louis Yang$^3$\footnote{E-mail address: Louis.Yang@physics.ucla.edu}
}
\affiliation{
$^1$Department of Physics, National Tsing Hua University, Hsinchu, Taiwan 300
\\
$^2$Physics Division, National Center for Theoretical Sciences, Hsinchu, Taiwan 300
\\
$^3$Department of Physics and Astronomy, University of California,
Los Angeles, California 90095, USA
}

\date{\today}
\begin{abstract}
The curvature singularity  in viable $f(R)$ gravity models is examined when the background density is dense.
This singularity could be eliminated  by adding the $R^{2}$ term in the Lagrangian.
Some of cosmological consequences, in particular the source for the scalar mode of gravitational waves, are discussed.
\end{abstract}


\maketitle

\section{Introduction}
\label{sec:introduction}

The accelerating expansion of the universe has been established by several cosmological observations, such as those from type Ia supernovae~\cite{astro-ph/9805201,astro-ph/9812133}, cosmic microwave background radiation~\cite{astro-ph/0302209,arXiv:1001.4538}, large scale structure~\cite{astro-ph/0501171} and weak lensing~\cite{astro-ph/0306046}. There are two ways to explain this phenomenon.
One is to add dark energy to modify matter and the other one is to modify gravity in the Einstein's equation.
The simplest version for the latter is
$f(R)$ gravity~\cite{XX1,XX2,arXiv:1002.4928}, which is by extending the Ricci scalar of $R$ to a function of $f(R)$ in the
Einstein-Hilbert action. As a result,
the late time accelerating universe can be realized in $f(R)$ gravity. Many viable $f(R)$ gravity models have been constructed by satisfying various conditions as well as constraints from cosmological observations~\cite{arXiv:1002.4928}.

The finite-time singularity problems~\cite{MG} have been examined in many modified theories,
such as $f(R)$ models~\cite{MGR}, modified Gauss-Bonnet models~\cite{MGG}, $f(T)$ models~\cite{MGT}, modified Horava-Lifshitz gravity~\cite{MGHL}
and non-local gravity models~\cite{MGLG}.
Recently, it has been pointed out in Refs.~\cite{S1, S2} that
  some of the viable $f(R)$ models contain one kind of the finite-time singularities, leading to a divergence of curvature, but it
  can be avoided by taking a fine-tune initial condition~\cite{Dev:2008rx, Appleby:2008tv}. However, this kind of singularities must be induced and can not be avoided when the local background density of matter becomes dense \cite{ arXiv:1012.1963}.
 Even though  the singularity
  depends on the background density as well as  the model parameters,  it happens in a finite time.
  This behavior could exist in many physical systems, such as cluster, galaxy, nebula collide, and star collapse.
 However,  if an additional  $R^{n}$ term with $ 1 < n \leq 2 $ is introduced in the viable $f(R)$ models~\cite{S3, Capozziello:2009hc,arXiv:1101.2820},
the singularity can be avoided.
We note that adding the $R^2$ term to the viable $f(R)$ gravity models could also lead to the unification of dark energy with inflation~\cite{XX1}.
We also remark that the curvature singularity arises naturally in the viable $f(R)$ models unless some fine-tuning is taken~\cite{arXiv:1101.2820}.

In this paper, we first show the singularity problem in the popular viable $f(R)$ models and then
try to modify them to remove the singularity.  We will explore the possible  cosmological consequences in these modified models
under the Minkowski background.
In particular, we will analyze
the scalar mode of gravitational waves~\cite{astro-ph/0307338,arXiv:1106.5582}, which is a characteristic signature to
distinguish  $f(R)$ gravity from general relativity (GR).
We will also show that the Minkowski approach still holds when the curvature close to the singularity.
We use the natural unit $c = \hbar = 1$
with $M_{pl}=G^{-1/2} \simeq 1.2 \times 10^{19} GeV$ and the metric  $g_{\mu \nu}=diag(-,+,+,+)$.

The paper is organized as follows. In Sec.~\ref{sec:oscillation},
we study the curvature oscillation in the viable $f(R)$ gravity models with a highly dense background.
In Sec.~\ref{sec:prevention}, we include the $R^{2}$ term in the models to prevent the singularity problem.
The scalar mode of gravitational waves is also discussed. The conclusions are given in Sec.~\ref{sec:conclusions}.

\section{Curvature Oscillation in $f(R)$ Gravity with Highly Dense Background\label{sec:oscillation}}
The action of $f(R)$ gravity is given by
\begin{equation}
S=\frac{1}{2\kappa^{2}}\int d^{4}x\sqrt{-g}(R+f(R))+S_{m}(g_{\mu\nu},\Upsilon_{\mu\nu})\,,
\label{eq:action}
\end{equation}
where $g$ is the determinant of the metric tensor $g_{\mu \nu}$, $S_{m}$ is the action of matter, $\kappa^2 \equiv 8\pi G =M_{pl}^{-2}$, $\Upsilon$ denotes the matter field with the minimal coupling to gravity, and $f(R)$ is an arbitrary function of the Ricci scalar $R$.
By varying the action (\ref{eq:action}) with respect to $g_{\mu \nu}$, we obtain equation of motion
\begin{equation}
(1+f_{R})R_{\mu \nu}-\frac{1}{2}(R+f)g_{\mu \nu}+(g_{\mu \nu} \square - \nabla_{\mu} \nabla_{\nu})f = \kappa^2 T_{\mu \nu},
\label{eq:EoM}
\end{equation}
where $f=f(R)$, the subscript $R$ denotes the derivative with respect to $R$, $i.e.$, $f_R = \partial f/\partial R$, $\nabla_{\mu}$ is the covariant derivative, $\square = g^{\mu \nu} \nabla_{\mu} \nabla_{\nu}$ is the d'Alembertian operator, and $T_{\mu \nu}$ is the matter energy-momentum tensor from the local matter distribution.

\subsection{Viable $f(R)$ Models}

It has been widely accepted that a viable $f(R)$ gravity model has to satisfy the following conditions~\cite{arXiv:1002.4928}: (a) $1+f_{R}> 0$ for $R>R_{0}$, which keeps the positivity of the effective gravitational coupling and avoids anti-gravity, where $R_{0}$ is the present background curvature; (b) $f_{RR}>0$ for $R>R_{0}$, which gives the stability condition of cosmological perturbations; (c) $f(R) \rightarrow R-2 \Lambda $ in the large curvature regime ($R \gg R_{0}$), which realizes the $\Lambda CDM$ behavior at $R \gg R_{0}$; (d) a stable late-time de-Sitter point; and (e) passing local gravity constraints, including those from
the equivalence principle and solar system. Under these conditions, many viable $f(R)$ models have been proposed \cite{arXiv:1002.4928}.
In Table \ref{table:models}, we give the explicit forms
 of the popular viable $f(R)$ models in the literature~\cite{arXiv:1002.4928}, where (i), (ii), (iii), (iv) and (v) correspond to Hu-Sawicki \cite{arXiv:0705.1158}, Starobinsky \cite{arXiv:0706.2041}, Tsujikawa \cite{arXiv:0709.1391}, the exponential gravity \cite{astro-ph/0511218, arXiv:0712.4017, arXiv:0905.2962, arXiv:1005.4574, Bamba:2011dk} and Appleby-Battye (AB) \cite{arXiv:0705.3199, arXiv:0909.1737} models, respectively.

\begin{table}[htbp]
\caption{Explicit forms of $f(R)$ in (i) Hu-Sawicki, (ii) Starobinsky, (iii) Tsujikawa, (iv) the exponential gravity, and (v) AB viable models}
\vskip 0.2in
\label{table:models}
\begin{tabular}{|c||c|c|} \hline
 model & $f(R)$ & Constant parameters
\\ \hline \hline
(i) & $ - R_{\mathrm{HS}} \frac{c_1  \left(R/R_{\mathrm{HS}}\right)^p}{c_2
\left(R/R_{\mathrm{HS}}\right)^p + 1}$ &
$c_1$, $c_2$, $p(>0)$, $R_{\mathrm{HS}}(>0)$
\\ \hline
(ii) & $ -\lambda R_{\mathrm{S}} \left[ 1- \left(1+\frac{R^2}{R_{\mathrm{S}}^2} \right)^{-n} \right]$ & $\lambda (>0)$, $n (>0)$, $R_{\mathrm{S}}$
\\ \hline
(iii) & $ - \mu R_{\mathrm{T}} \tanh\left( \frac{R}{R_{\mathrm{T}}} \right)$ & $\mu (>0)$, $R_{\mathrm{T}} (>0)$
\\ \hline
(iv) & $ -\beta R_{\mathrm{E}}\left(1-e^{-R/R_{\mathrm{E}}}
\right)$ & $\beta$, $R_{\mathrm{E}}$
\\ \hline
(v) &  $-g R + g R_{AB} \ln \left[ \frac{\cosh\,\left(R/R_{\mathrm{AB}}-b\right)}{\cosh\,b}\right]$
& $g$, $b$, $R_{\mathrm{AB}}$
\\ \hline
\end{tabular}
\end{table}

\subsection{Curvature Oscillation in $f(R)$ Gravity}

We start from the trace of the field equation in Eq.~(\ref{eq:EoM}), given by
\begin{equation}
\label{eq:EoMtrace}
Rf_{R}-2f-R+3\square f_{R}=\kappa^{2}T,
\end{equation}
where $T=g^{\mu \nu}T_{\mu \nu}$ is the trace of the energy-momentum tensor. This equation is reduced to GR
 ($R=-\kappa^{2} T$) if $f(R)=0$. Clearly, it contains an extra degree of freedom beyond GR when $f_{R}\neq 0$.
 Note that Eq.(\ref{eq:EoMtrace}) is a fourth order field equation in comparison with the second order one in GR. This fourth order equation also leads to the oscillation behavior of the Ricci scalar.

Before calculating the curvature oscillation  in the viable $f(R)$ models, we briefly introduce  the framework of our study.
It is known that the curvature in the dense matter regime, such as inner-galaxy, nebula and star collapse,
 is much bigger than the background curvature of the universe. Note that the average density of the universe
and  the density inside the inner galaxy are
  $10^{-29}$ and $10^{-24} g/cm^{3}$, respectively. In the large curvature limit, the viable $f(R)$ models in Table\ref{table:models} can be reduced
  into power law and exponential types, given by
\begin{eqnarray}
\label{eq:HSSType}
&(I): &  \quad  f(R) \simeq -\lambda R_{ch}\left[1-\left(\frac{R_{ch}}{R}\right)^{2n}\right],\\
\label{eq:ExpType}
&(II): &  \quad  f(R) \simeq -\lambda R_{ch}\left(1-e^{-R/R_{ch}} \right),
\end{eqnarray}
referred to as Type-I and II,
respectively, where $R_{ch}$ is a dimension-two constant.
The Hu-Sawicki and Starobinsky models belong to Type-I in Eq.~(\ref{eq:HSSType}), while Tsujikawa, AB and exponential models correspond to
Type-II in Eq.~(\ref{eq:ExpType}). In order to explain the accelerating universe, the  parameter $R_{ch}$ should be the same order as
the background value of $R_{0}$ at the current epoch.

\subsubsection{Curvature Oscillation  in  Type-I $f(R)$ Gravity}
We consider $f(R)$ gravity in dense, locally homogeneous and isotropic perfect fluid with the non-relativistic matter density $\rho_{m}$, and assume that the density changes homogeneously in time and is much denser than the critical density $\rho_{c}$. Then, the trace of the energy-momentum tensor can be expressed as
\begin{equation}
\label{eq:energy-momentum}
T=-T_{0}\left(1+\frac{t}{t_{ch}} \right),
\end{equation}
where
$T_{0}=\rho_{m}^{(0)}-3P_{m}^{(0)} \simeq \rho_{m}^{(0)}$
is the initial background density and $t_{ch}$ is the characteristic time. By defining a dimensionless variable
\begin{equation}
\label{eq:variable}
u=R_{ch}/R\,,
\end{equation}
$f$ and $f_{R}$ of Type-I  in Eq.(\ref{eq:HSSType}) become
\begin{eqnarray}
\label{eq:Osc01_1}
&f& \simeq -\lambda R_{ch} \left[1-\left(\frac{R_{ch}}{R} \right)^{2n}\right]=-\lambda R_{ch} \left(1-u^{2n}\right), \\
\label{eq:Osc01_2}
&f_{R}& \simeq -2n \lambda \left(\frac{R_{ch}}{R}\right)^{2n+1}=-2n \lambda u^{2n+1},
\end{eqnarray}
respectively. The trace of the field equation in Eq.~(\ref{eq:EoMtrace}) results in the oscillation behavior in the large density limit, given by
\begin{eqnarray}
\label{eq:Osc02}
\square f_{R} &=& 2n\lambda \frac{d^{2}}{d t^{2}}\left(u^{2n+1}\right) = 2n\left( 2n+1 \right)\lambda \left( u^{2n} \ddot{u} + 2n u^{2n-1} \dot{u}^{2}\right) \nonumber \\
&=& -\frac{1}{3} \left\{ \kappa^{2}T_{0} \left( 1 + \frac{t}{t_{0}} \right) -R_{ch}\left[ 2n\lambda u^{2n} -2\lambda \left(1 - u^{2n} \right) +u^{-1} \right] \,,\right\}
\end{eqnarray}
where we have assumed that the curvature and energy density depend on time only. By using the dense background
$\kappa^{2} T_{0}/R_{ch} \simeq R/R_{ch} (=u^{-1}) \sim \rho_{m} / \rho_{c} \gg 1$, Eq.~(\ref{eq:Osc02}) can be simplified to
\begin{eqnarray}
\label{eq:Osc03}
\ddot{u} + 2n \frac{\dot{u}^{2}}{u} + \frac{u^{-2n}}{2n(2n+1) \lambda} \left[ \frac{\kappa^{2} T_{0}}{3} \left( 1 + \frac{t}{t_{ch}} \right) - \frac{R_{ch}}{3u} \right] = 0\,.
\end{eqnarray}
With the rescaled  variables $y=\beta u$ and $\tau =\gamma^{-1} t$, we obtain
\begin{eqnarray}
\label{eq:Osc04}
y^{\prime\prime} + 2n \frac{y^{\prime 2}}{y}+ y^{-2n} \left[ \left( 1+ \frac{\tau}{\tau_{ch}} \right) -y^{-1} \right] = 0,
\end{eqnarray}
where
\begin{eqnarray}
\label{eq:Osc05}
\beta &=& \frac{\kappa^{2} T_{0}}{R_{ch}},  \\
\gamma^{2} &=& \frac{6n\left( 2n+1 \right) \lambda }{R_{ch}} \left( \frac{R_{ch}}{\kappa^{2} T_{0}} \right)^{2n+2},
\end{eqnarray}
and  the prime denotes the derivative with respect to $\tau$.
Note that $\beta$ is a dimensionless parameter, which rescales $y=\beta u= \kappa^{2} T_{0}/R$ to be unity when the background density is stationary at the initial value of $R \simeq \kappa^{2} T_{0}$, while $\tau$ is also a dimensionless variable, related to the physical time by a constant and time dimension factor $\gamma$. Since $t_{ch} = \gamma \tau_{ch}$,
 $\gamma $ can be estimated under various backgrounds and model parameters. For example, $\gamma \sim (400,4 \times 10^{-3})$s with $n=(2,3)$ and $\rho_{m} \simeq 10^{-24} g/cm^{3}$.

\begin{center}
\begin{figure}[tbp]
\begin{tabular}{ll}
\begin{minipage}{80mm}
\begin{center}
\unitlength=1mm
\resizebox{!}{6.5cm}{\includegraphics{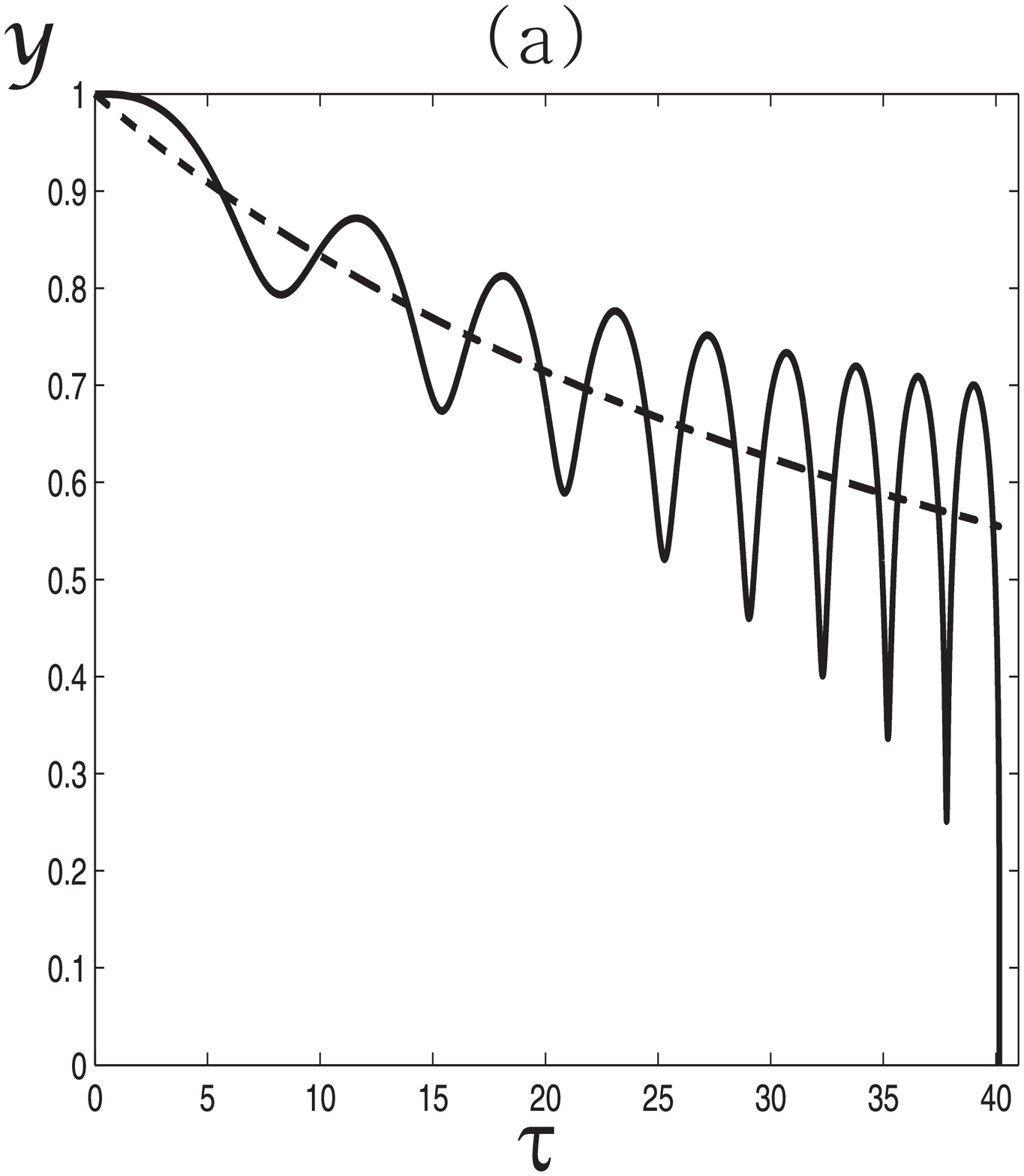}}
\end{center}
\end{minipage}
&
\begin{minipage}{80mm}
\begin{center}
\unitlength=1mm
\resizebox{!}{6.5cm}{\includegraphics{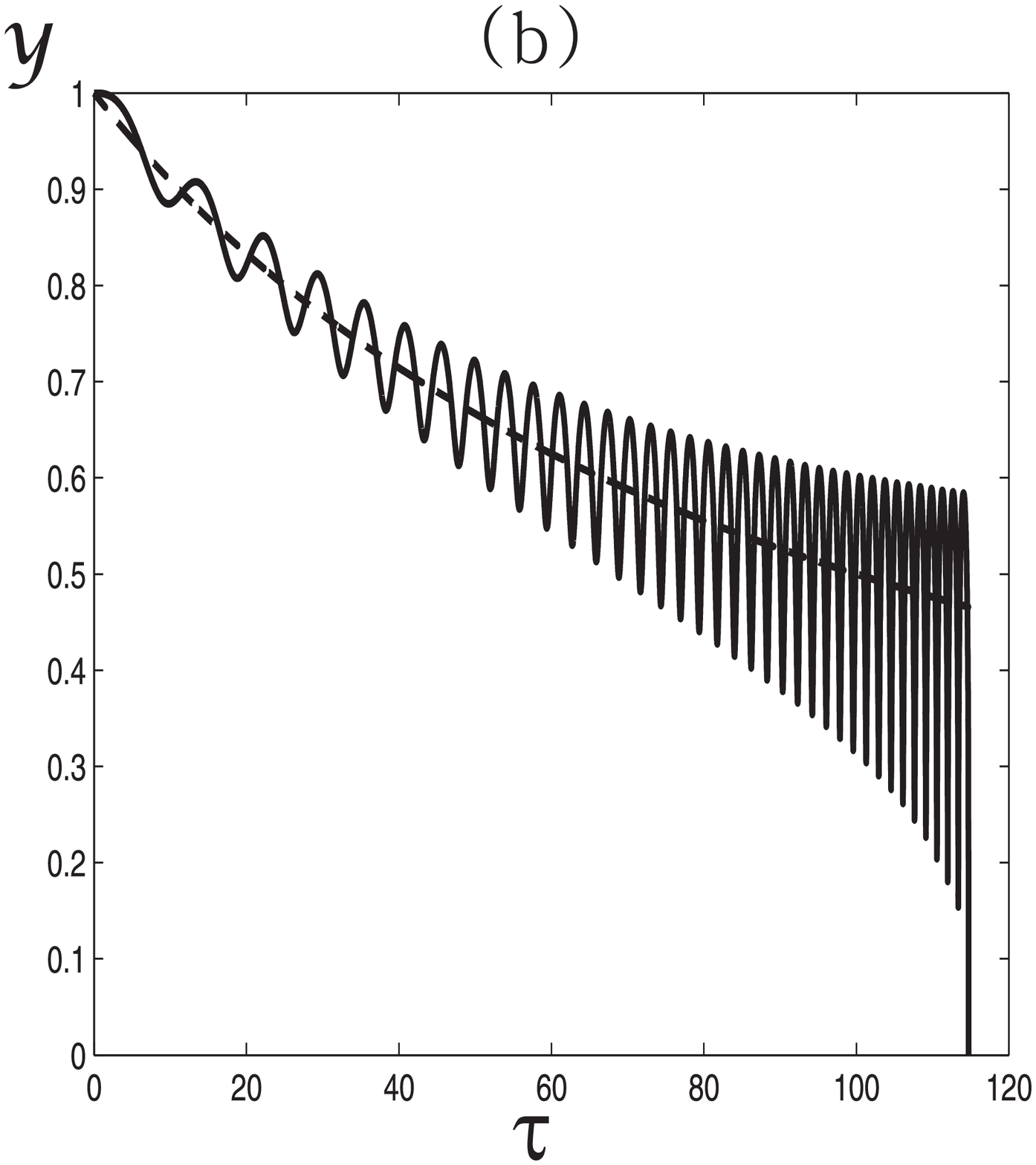}}
\end{center}
\end{minipage}\\[5mm]
\end{tabular}
\caption{Evolution of $y=\kappa^{2}T_{0}/R$ as functions of the rescaled time $\tau$ in the Type-I models with n=2 and (a) $\tau_{ch}=50$
and (b) $\tau_{ch}=100$, where
the dashed lines correspond to $R=-\kappa^{2} T= \kappa^{2} T_{0}(1+\tau/ \tau_{ch})$.}
\label{fg:1}
\end{figure}
\end{center}

In Fig.~\ref{fg:1}, we show the evolution of $y=\kappa^{2}T_{0}/R$ as functions of the rescaled time $\tau$ in the Type-I models
 with initial conditions of  $ y_{0}=1 $ and $ y^{\prime}_{0} =0 $.
 From the figures, we see that  the curvature singularity of  $R \rightarrow \infty$ appears   when $y= \kappa^2 T_{0} / R$ reaches zero.
 Clearly, the singularity exists in a time shorter than the age of the universe.
 The evolution equation (\ref{eq:Osc04}) only depends on the characteristic time $\tau_{ch}$, so that the divergence always happens at the order of $\tau_{ch}$.

\subsubsection{Curvature Oscillation  in  Type-II $f(R)$ Gravity}

\begin{center}
\begin{figure}[tbp]
\begin{tabular}{ll}
\begin{minipage}{80mm}
\begin{center}
\unitlength=1mm
\resizebox{!}{6.5cm}{\includegraphics{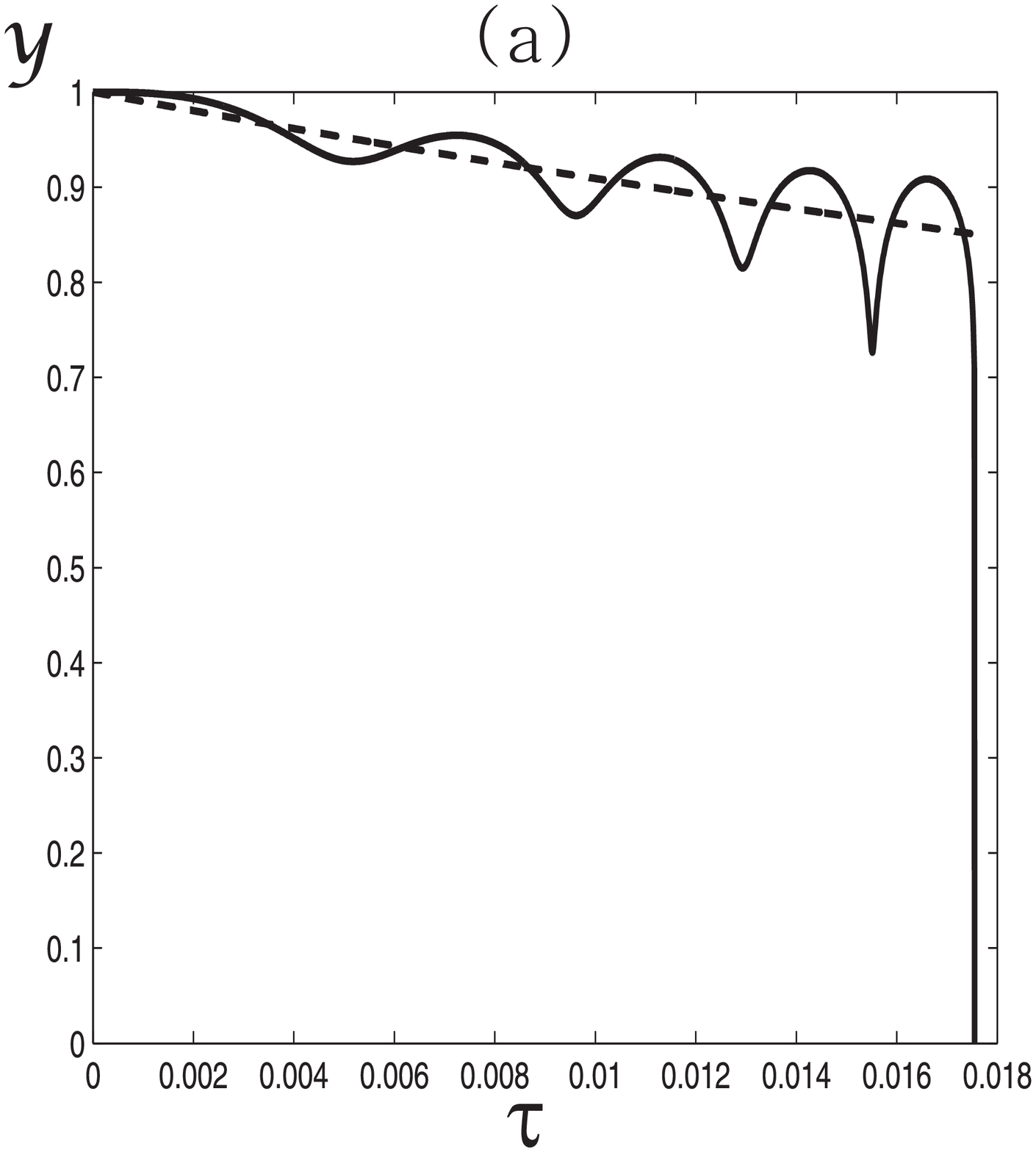}}
\end{center}
\end{minipage}
&
\begin{minipage}{80mm}
\begin{center}
\unitlength=1mm
\resizebox{!}{6.5cm}{\includegraphics{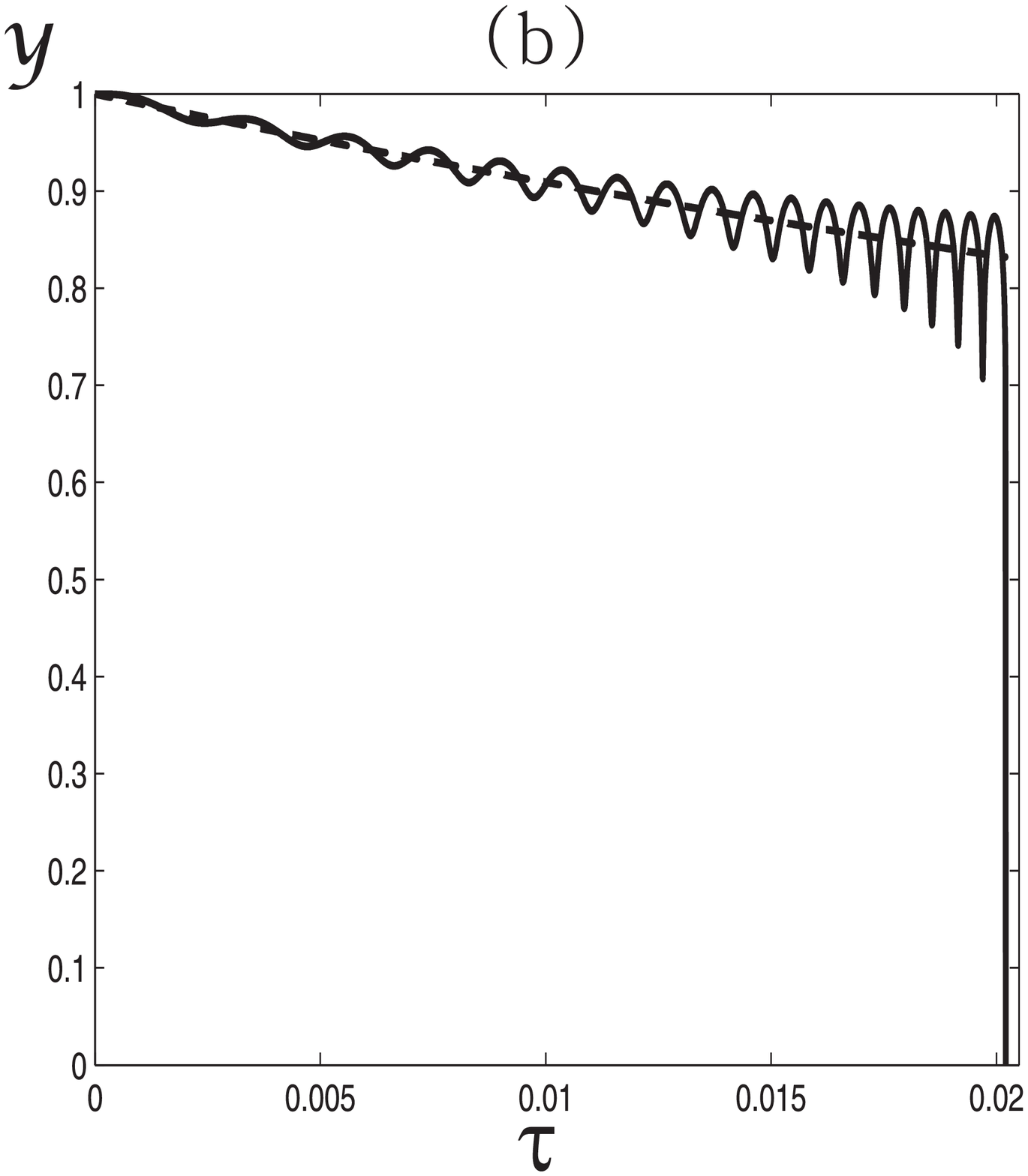}}
\end{center}
\end{minipage}\\[5mm]
\end{tabular}
\caption{Evolution of $y=\kappa^{2}T_{0}/R$
 as functions of the rescaled time $\tau$ in the Type-II models with
$\tau_{ch}=0.1$ and (a) $\beta=16$ and (b) $\beta=18$,
 where the dashed lines correspond to $R=-\kappa^{2} T= \kappa^{2} T_{0}(1+\tau/ \tau_{ch})$.}
\label{fg:2}
\end{figure}
\end{center}

We now consider the divergent behavior in the Type-II viable $f(R)$ models. In the dense background density, from Eq.~(\ref{eq:ExpType})
 we have
\begin{eqnarray}
\label{eq:Osc11}
f_{R}& \simeq &-\lambda e^{-R/R_{ch}},
\end{eqnarray}
Substituting Eqs.~(\ref{eq:ExpType}) and (\ref{eq:Osc11})
 into the trace equation (\ref{eq:EoMtrace}), we obtain
\begin{eqnarray}
\label{eq:Osc12}
\ddot{x}-\dot{x}^{2}+\frac{x R_{ch}}{3 \lambda} e^{x} - \frac{\kappa^2 T_{0}}{3 \lambda} e^{x} \left(1+ \frac{t}{t_{ch}} \right) \simeq 0,
\end{eqnarray}
where $x\equiv u^{-1}=R/R_{ch}$. Because  the e-folding contains a variable $x$, it is hard to rescale this equation into a background independent equation
as that in Eq.~(\ref{eq:Osc04}). However, we can still redefine some parameters to modify the evolution equation to
\begin{eqnarray}
\label{eq:Osc13}
y^{\prime \prime}+(\beta y^{-1} - 2 )\frac{y^{\prime 2}}{y}+ y^{2}
 e^{\beta/y }
 \left[ \left(1+\frac{\tau}{\tau_{ch}}\right) -y^{-1} \right]=0,
\end{eqnarray}
where $\beta = \kappa^{2} T_{0} / R_{ch}$, $y \equiv \beta x^{-1}= \kappa^{2} T_{0} / R$ and the prime denotes the derivative with respect to $\tau$, defined by $\tau=\xi^{-1} t$ with
$\xi^2=3\lambda / R_{ch} $.
The initial conditions can be determined easily to be $y_{0}=1$ and $y^{\prime}_{0}=0$ when the background is stationary.
The time scaling factor can be estimated as $\xi \sim 4.3 \times 10^{17} $s.
In Fig.~\ref{fg:2}, we illustrate the evolution of $y$ as functions of $\tau$ in the Type-II viable $f(R)$ models. The figures show
 that the divergent behavior  depends very strongly on the background, and the singularity appears not only in a finite, but a very short time.

\section{Preventing The Singularity Problem}\label{sec:prevention}
\subsection{ The $R^{2}$ Term in $f(R)$ Gravity}
It has been shown that the singularity could be prevented by an additional $R^{m} / M^{ 2\left( m-1 \right) }$ term with $1< m \leq 2$~\cite{arXiv:1101.2820}. Since the
inflationary \cite{157549, HUTP-85-A017,CALT-68-1320,Hwang:2011kg} and evolutionary~\cite{Arbuzova:2011fu} model
of $f(R)=R + R^{2} / M^{2}$
 has been well-discussed, it is reasonable to examine its behavior with $m=2$. In the large curvature regime, we rewrite the viable $f(R)$ models
in Eqs.~(\ref{eq:HSSType}) and (\ref{eq:ExpType})
plus the $R^{2}$ term as
\begin{eqnarray}
\label{eq:PrevHSSType}
&(I):& \quad f(R) = f_{HSS} + f_{R^{2}} \simeq -\lambda R_{ch}\left[1-\left(\frac{R_{ch}}{R}\right)^{2n}\right] + \frac{R^{2}}{M^{2}} , \nonumber \\
\label{eq:PrevExpType}
&(II):& \quad f(R) = f_{Exp} + f_{R^{2}} \simeq -\lambda R_{ch}\left(1-e^{-R/R_{ch}} \right) + \frac{R^{2}}{M^{2}}\,,
\end{eqnarray}
respectively.

\begin{center}
\begin{figure}[tbp]
\begin{tabular}{ll}
\begin{minipage}{80mm}
\begin{center}
\unitlength=1mm
\resizebox{!}{6.5cm}{\includegraphics{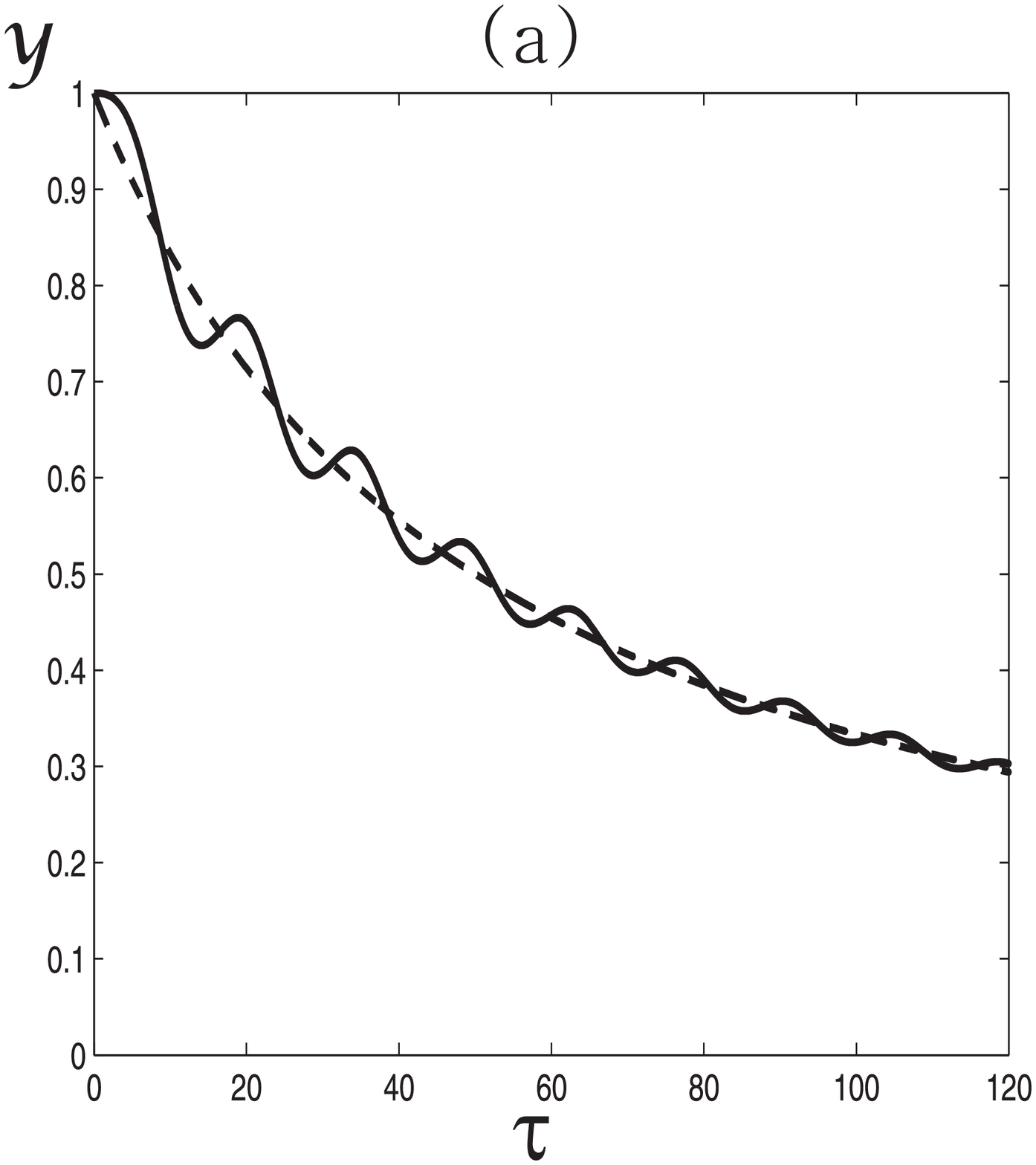}}
\end{center}
\end{minipage}
&
\begin{minipage}{80mm}
\begin{center}
\unitlength=1mm
\resizebox{!}{6.5cm}{\includegraphics{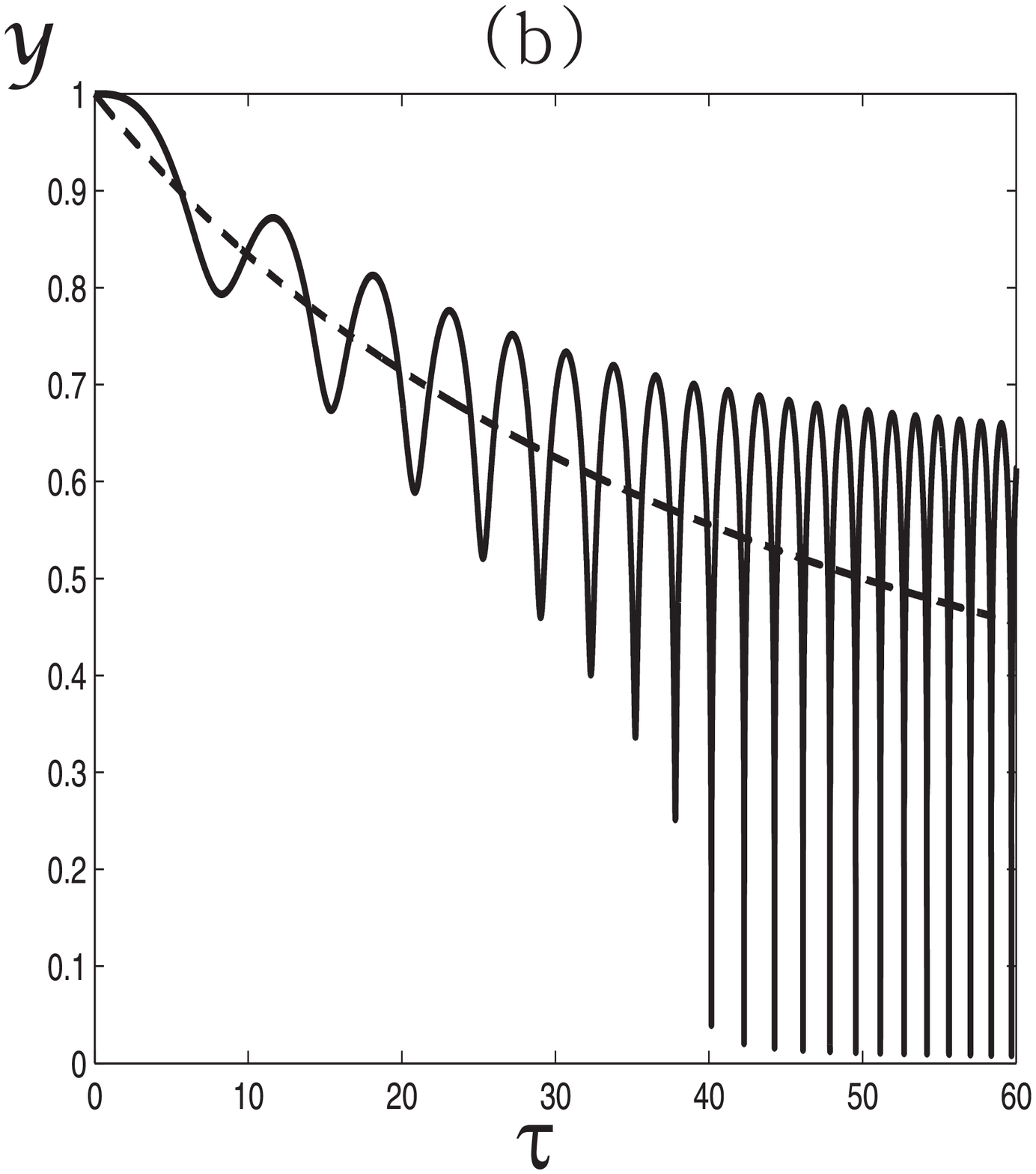}}
\end{center}
\end{minipage}\\[5mm]
\end{tabular}
\caption{Legend is the same as Fig.~\ref{fg:1} but with $\tau_{ch}=50$ and (a) $g_I=1$ and (b) $g_I=10^{-6}$.
}
\label{fg:3}
\end{figure}
\end{center}

We now examine whether the singularity problem can be resolved in the above modified viable $f(R)$ models.
In the modified Type-I models, with the similar procedure in Eq.~(\ref{eq:Osc03}), the trace equation (\ref{eq:EoMtrace}) can be rewritten
 as
\begin{eqnarray}
\label{eq:PrevPLNum}
&&y^{\prime \prime} + 2n \frac{y^{\prime 2} }{y} + g_{I} y^{-\left(2n+2\right)} \left( y^{\prime \prime} - \frac{2y^{\prime 2}}{y} \right) +y^{-2n} \left[ -y^{-1} + \left( 1+ \frac{\tau}{\tau_{ch}} \right) \right]=0 ,
\end{eqnarray}
where
$y=\kappa^{2} T_{0}/R$, $\tau=\gamma^{-1}t$ and $g_{I} $ arises from the addition $R^{2}$ term, given by
\begin{eqnarray}
\label{eq:PrevHSSPara}
g_{I} = \frac{R_{I}}{\lambda n (2n+1) M^{2}} \left( \frac{\kappa^{2} T_{0}}{R_{I}} \right)^{2n+2},
\end{eqnarray}
with the time rescaling factor $\gamma^{2}=(6 \lambda n (2n+1)/R_I) (R_I/(\kappa^2 T_0) )^{2n+2}$ and
 $R_{I} = R_{ch}$. In Eq.~(\ref{eq:PrevPLNum}), the curvature singularity could be prevented by the additional factor $g_{I} $
 since there is a huge restoration force in a large curvature regime.
 Similar to that in Eq.~(\ref{eq:Osc04}), one finds that $R \rightarrow \infty$ if $y \rightarrow 0$ in Eq.~(\ref{eq:PrevPLNum}). Moreover, Eq.~(\ref{eq:PrevPLNum}) is also a scale independent evolution equation with the period of the oscillation depending on the time scaling factor $\gamma$. We illustrate the oscillation behavior in Fig.~\ref{fg:3}. Clearly, the singularity can be avoided when we include
  the $R^{2} $ term in the Lagrangian. It still holds even when $g_{I} \ll 1$,  but the amplitude is strongly
  related to the dimensionless positive parameter $g_{I}$ in Eq.~(\ref{eq:PrevPLNum}).
  We note that the singularity appears at $\tau \simeq 40$ if there is no $R^{2}$ term.
  We also note that $R_{I}$ is determined from cosmological constraints, which should be the same order as the cosmological constant.

\begin{center}
\begin{figure}[tbp]
\begin{tabular}{ll}
\begin{minipage}{80mm}
\begin{center}
\unitlength=1mm
\resizebox{!}{6.5cm}{\includegraphics{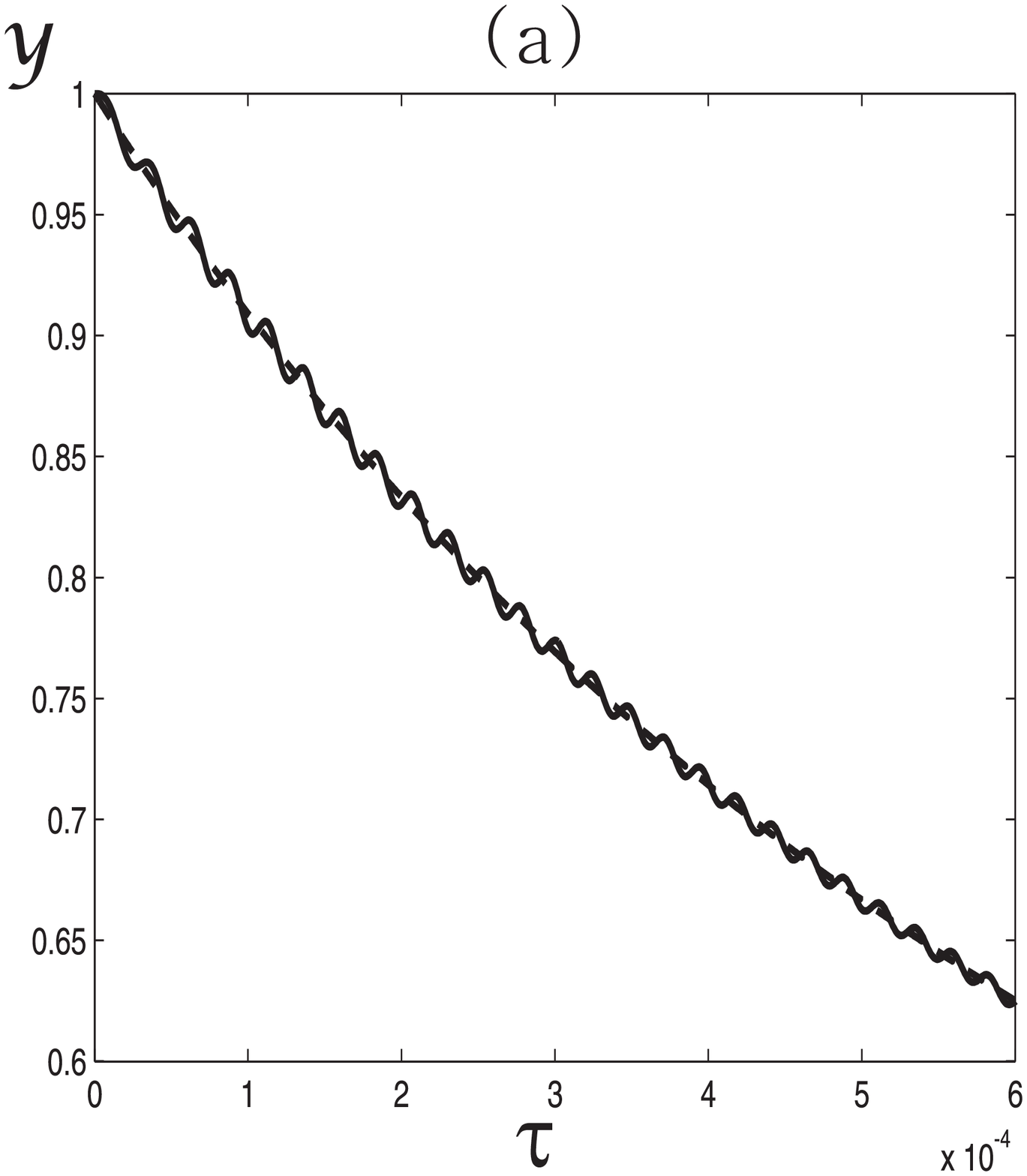}}
\end{center}
\end{minipage}
&
\begin{minipage}{80mm}
\begin{center}
\unitlength=1mm
\resizebox{!}{6.5cm}{\includegraphics{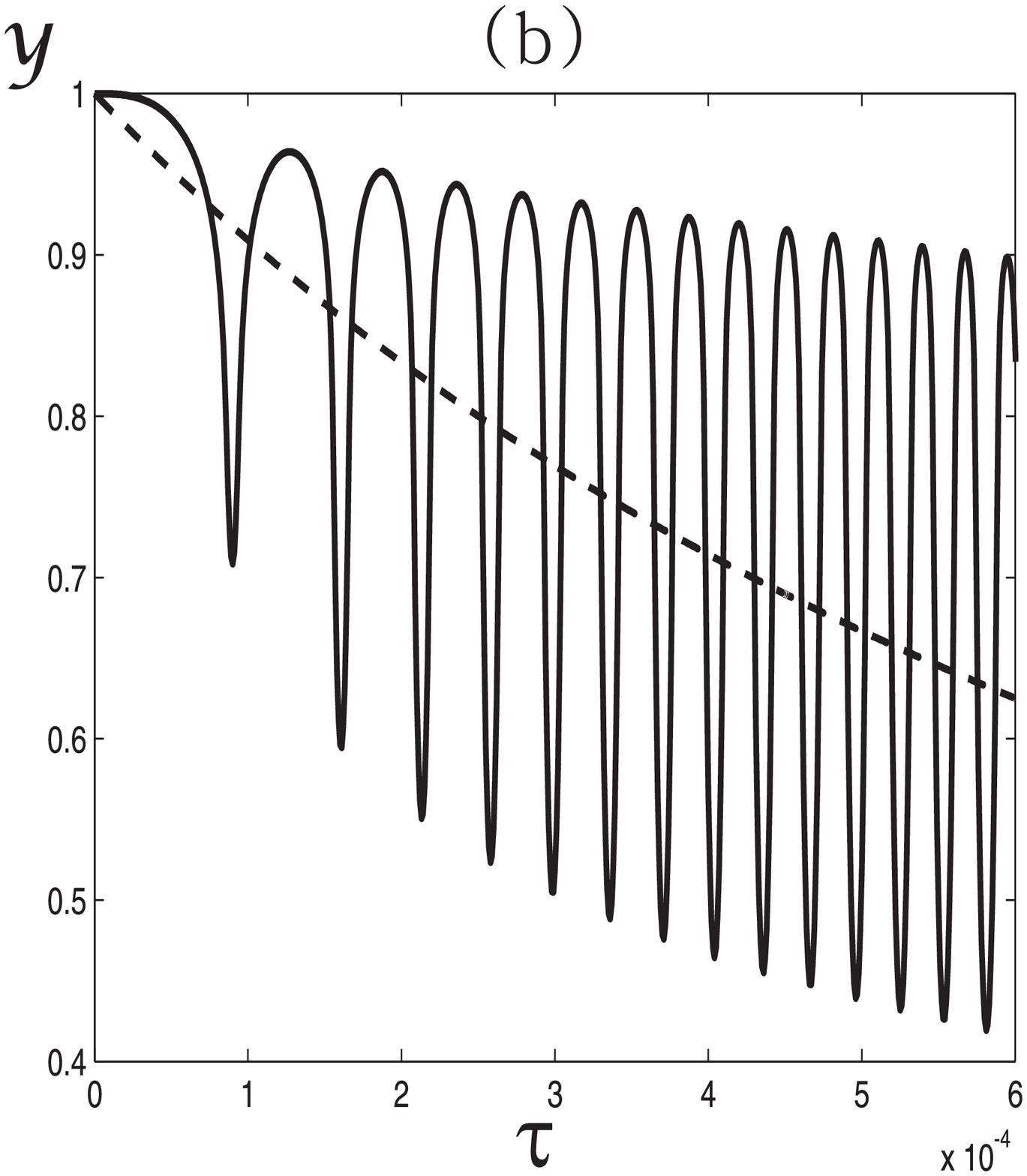}}
\end{center}
\end{minipage}\\[5mm]
\end{tabular}
\caption{
Legend is the same as Fig.~\ref{fg:2} but for
$\tau_{ch}=10^{-3}$
 and (a) $g_{II}=1$ with $\beta=25$ and (b)
$g_{II}=10^{-2}$
 with $\beta\simeq 20.4$ .
}
\label{fg:4}
\end{figure}
\end{center}

In the Type-II $f(R)$ models, the evolution equation is
\begin{eqnarray}
\label{eq:PrevExpNum}
&&y^{\prime \prime} + \left( \beta y^{-1} -2 \right) \frac{y^{\prime 2} }{y} + g_{II} e^{\beta\left( \frac{1}{y} -1 \right)} \left( y^{\prime \prime} - \frac{2y^{\prime 2}}{y} \right) + y^{2}
e^{\beta/y}
 \left[ \left( 1+ \frac{\tau}{\tau_{ch}} \right)  -y^{-1} \right]=0 ,
\end{eqnarray}
where
$y=\kappa^{2}T_{0}/R$, $\tau=\xi^{-1}t$,
$\xi^2=3 \lambda/R_{II}$
and
$g_{II} = \frac{2 R_{II}}{3 \lambda M^{2}}e^{\beta}$.
Similarly, as in the modified Type-I models, the singularity could be eliminated when the $R^{2}$ term is added.
In Fig.~\ref{fg:4},
we show the curvature oscillation with two different values of $g_{II}$.
Clearly, its amplitude of the oscillation still strongly depends on $g_{II}$.

 From the above results, one can easily conclude that the oscillation behavior is determined by the dimensionless constants $g_{I,II}$.
  Consequently, if the curvature is large enough, the curvature oscillates as a simple harmonic oscillator with a driving term ``$\kappa^{2} T$''.
  If the curvature is not large enough,
it is dominated by the original parts of the $f(R)$ models, whereas the singularity is removed by adding the $R^{2}$ term
in the large curvature regime.

\subsection{Cosmological Phenomena in Modified Viable Models with $R^{2}$}

Although we can avoid the singularity with introducing the $R^{2}$ term in the Lagrangian, the oscillation behavior still exists. This behavior might appear in some physical systems such as the scalar mode of gravitational waves,
which has been recently discussed in
Ref.~\cite{arXiv:1106.5582} for the viable $f(R)$ models. The graviton in GR is a spin-two massless particle with two spin polarizations, corresponding
``plus'' and ``cross" modes, respectively.
The  scalar mode  is an extra mode of gravitational waves,  coming from the additional degree of freedom in $f(R)$ gravity and the non-vanishing trace equation in vacuum (we can estimate it by using Eq.~(\ref{eq:EoMtrace}) with $T=0$).
 The scalar mode of gravitational waves propagates in the vacuum like a massive scalar field $\square h_{f} = m_{s}^2 h_{f}$, where $R_{min}$ is the background curvature, $m_{s}^{2}=\frac{1}{3} \left( \frac{ 1+ f_{R}(R_{min}) }{f_{RR}(R_{min})} - R_{min} \right)$ and $h_{f} \equiv  \frac{\delta f_{R}}{1+ f_{R}\left(R_{min}\right)} = \frac{ R_{min}f_{RR}\left( R_{min} \right)}{ 1+ f_{R}\left( R_{min} \right)} \frac{\delta R}{R_{min}}$. Because of the same origin, the density increasing system could be a source of the scalar mode of gravitational waves when the the curvature oscillation amplitude ($\delta R / R_{min}$) is large.
\begin{center}
\begin{figure}[tbp]
\begin{tabular}{ll}
\begin{minipage}{80mm}
\begin{center}
\unitlength=1mm
\resizebox{!}{6.5cm}{\includegraphics{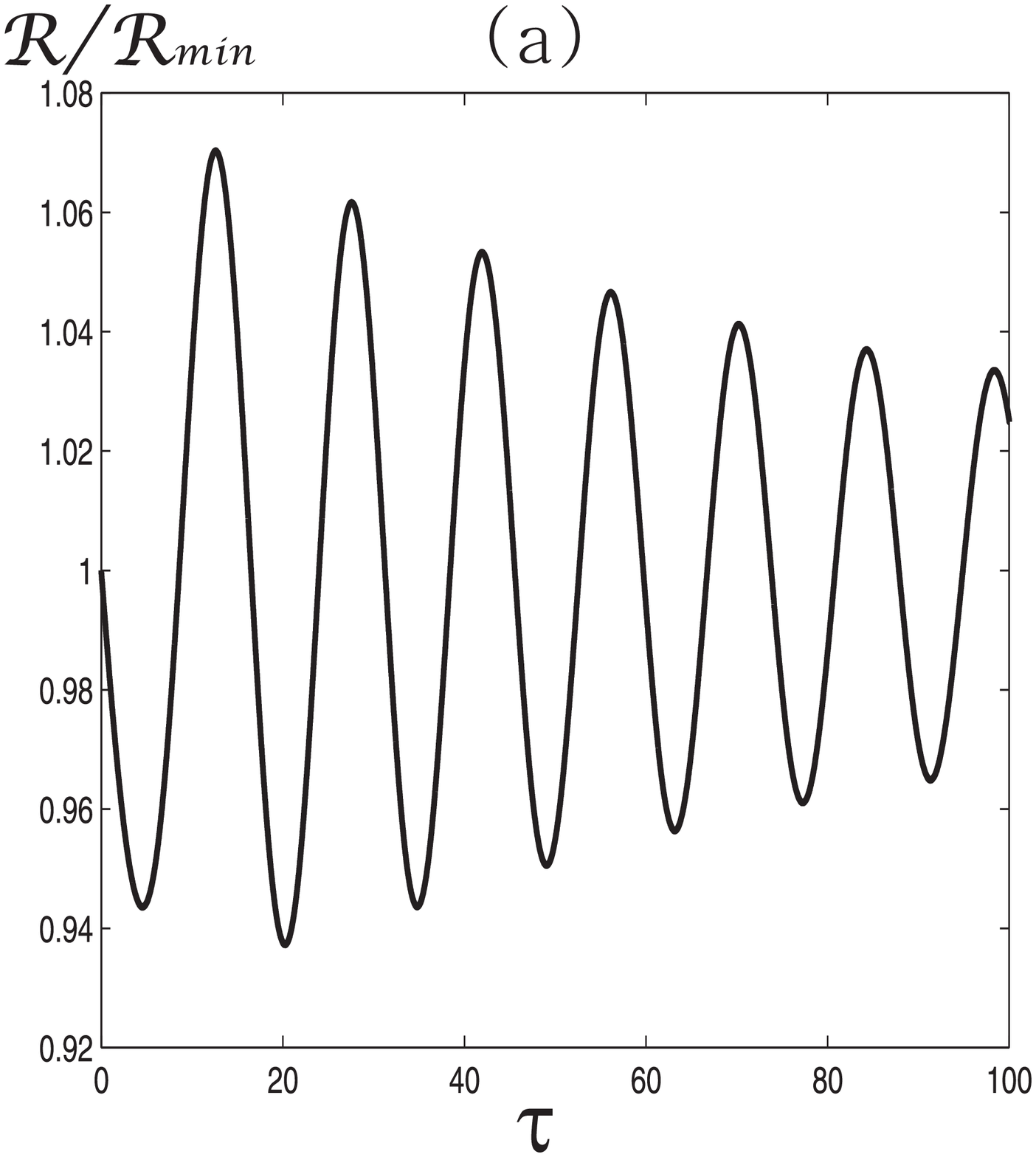}}
\end{center}
\end{minipage}
&
\begin{minipage}{80mm}
\begin{center}
\unitlength=1mm
\resizebox{!}{6.5cm}{\includegraphics{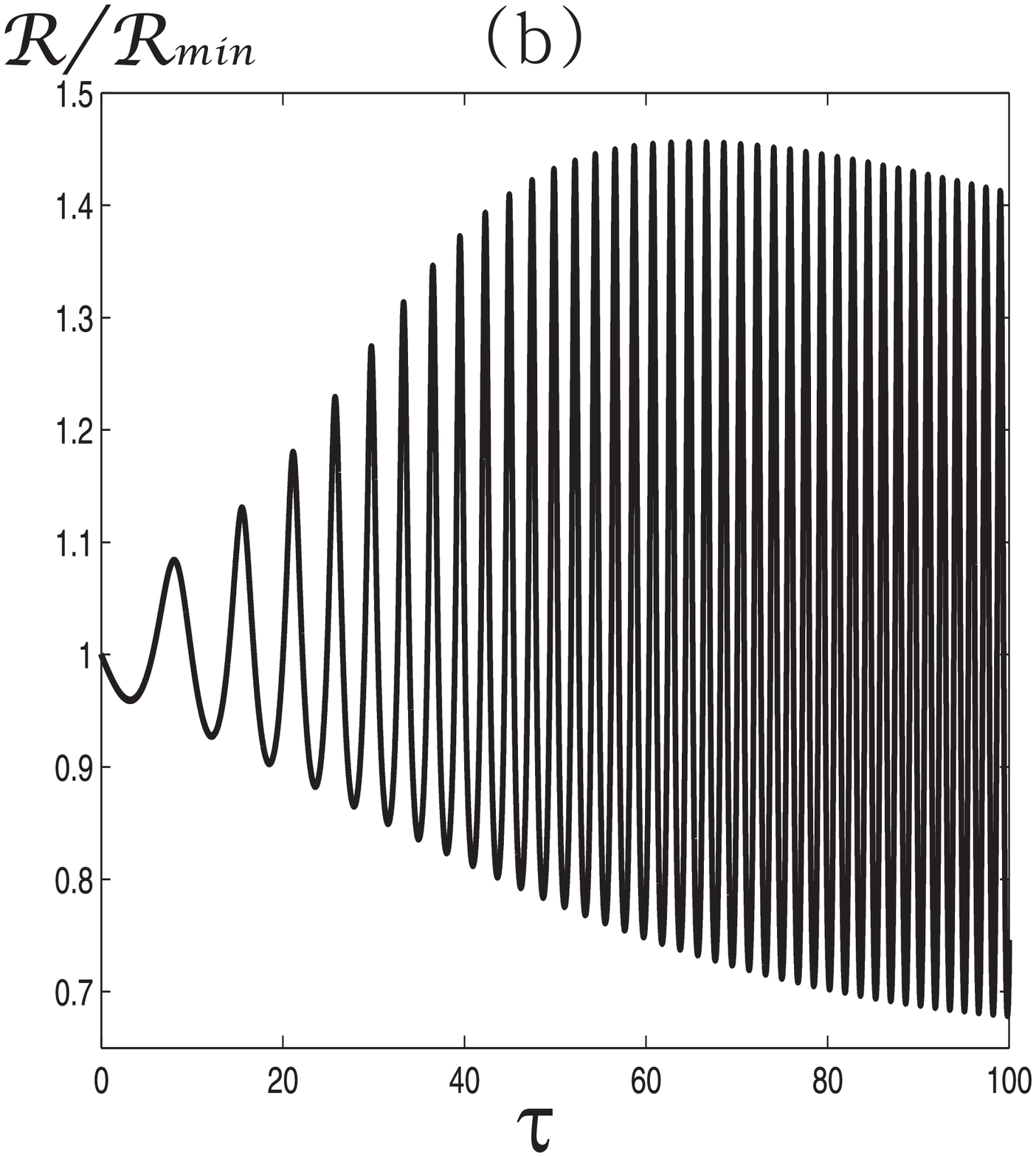}}
\end{center}
\end{minipage}\\[5mm]
\begin{minipage}{80mm}
\begin{center}
\unitlength=1mm
\resizebox{!}{6.5cm}{\includegraphics{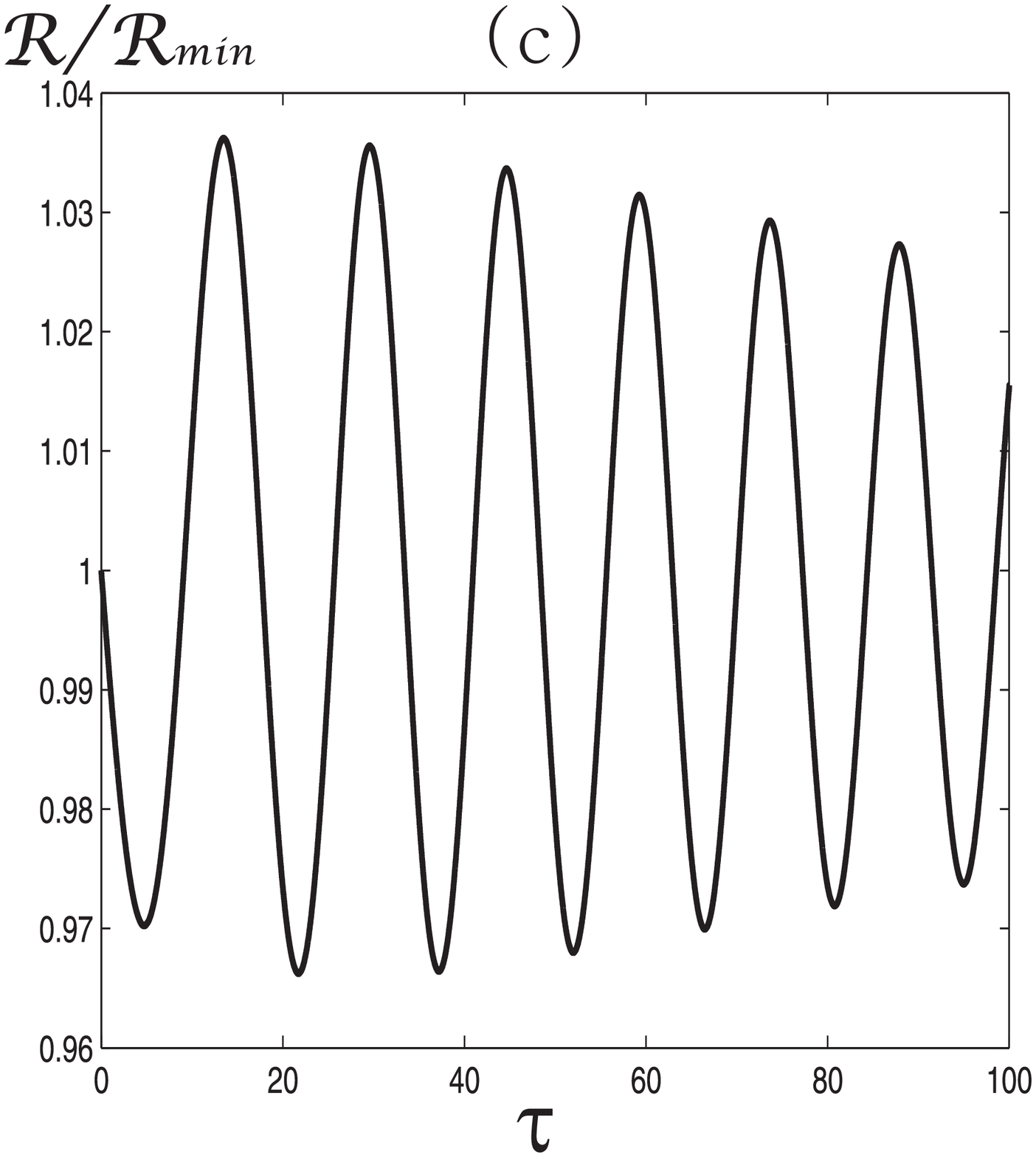}}
\end{center}
\end{minipage}
&
\begin{minipage}{80mm}
\begin{center}
\unitlength=1mm
\resizebox{!}{6.5cm}{\includegraphics{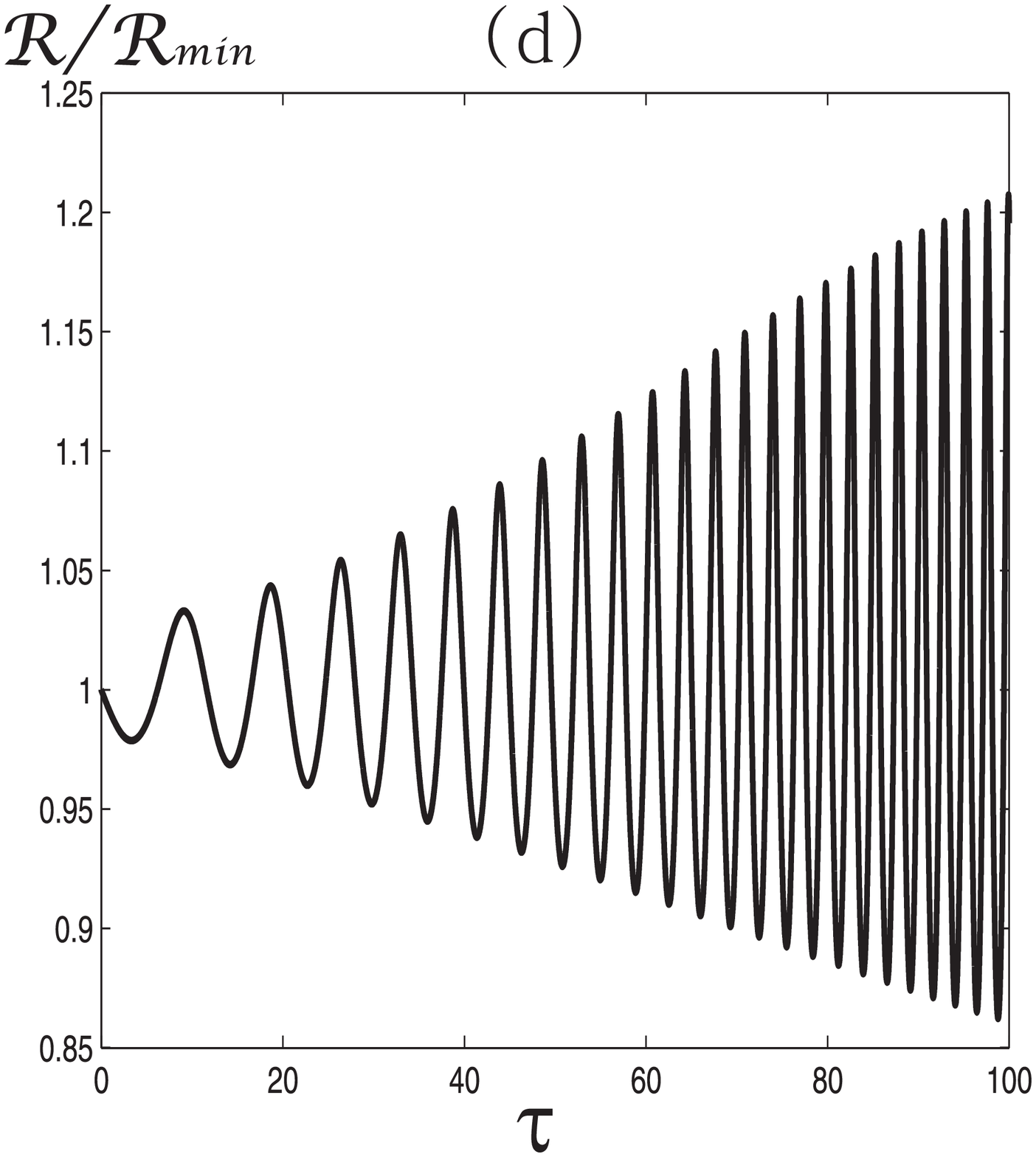}}
\end{center}
\end{minipage}
\end{tabular}
\caption{$R/R_{min}$ as functions of $\tau$ in the modified Type-I $f(R)$ models with $n=2$, where
(a), (b), (c), and (d) represent $(g_I,\tau)=(1,50)$, $(10^{-2},50)$, $(1,100)$, and $(10^{-2},100)$, respectively.
}
\label{fg:5}
\end{figure}
\end{center}

\begin{center}
\begin{figure}[tbp]
\begin{tabular}{ll}
\begin{minipage}{80mm}
\begin{center}
\unitlength=1mm
\resizebox{!}{6.5cm}{\includegraphics{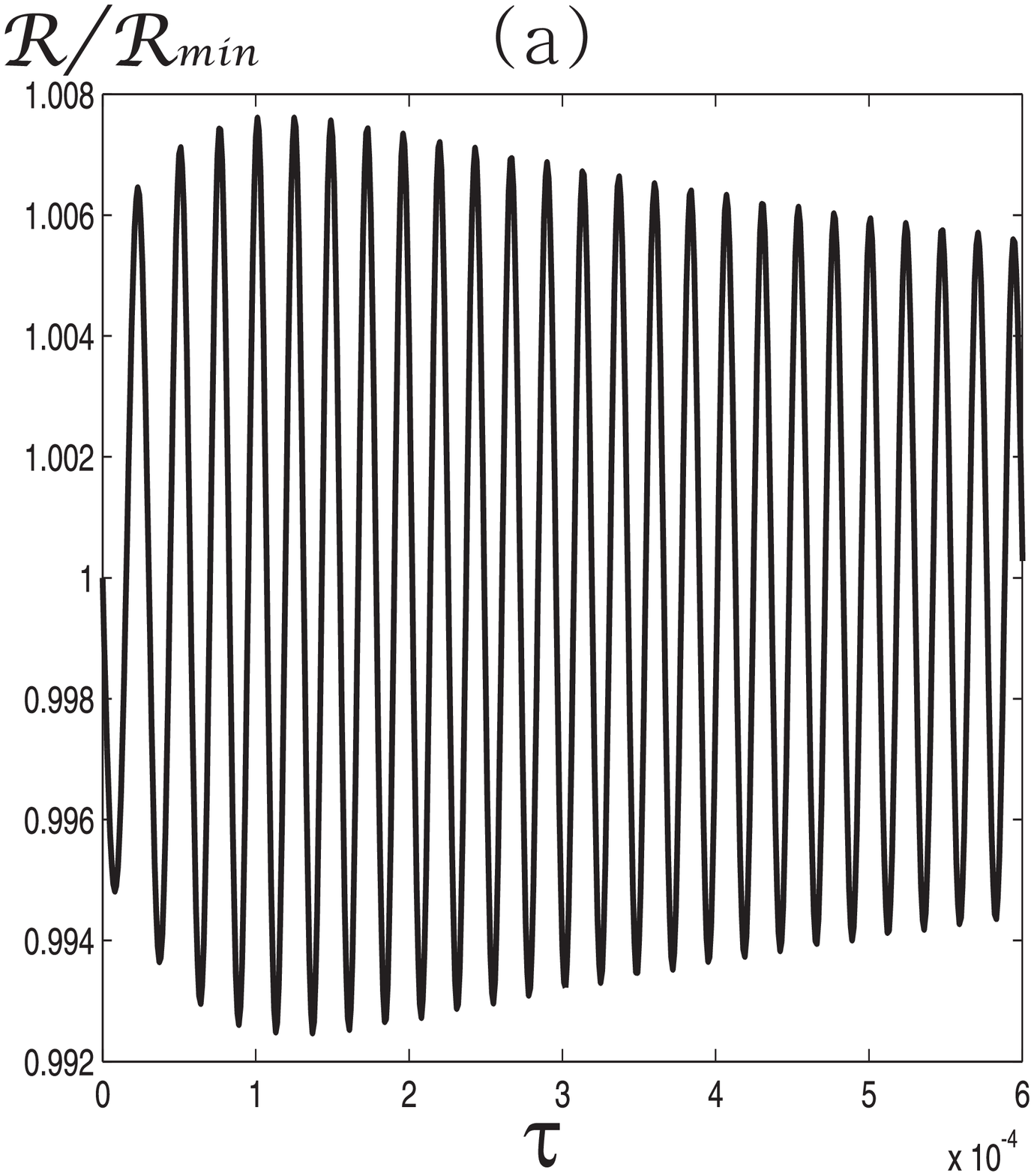}}
\end{center}
\end{minipage}
&
\begin{minipage}{80mm}
\begin{center}
\unitlength=1mm
\resizebox{!}{6.5cm}{\includegraphics{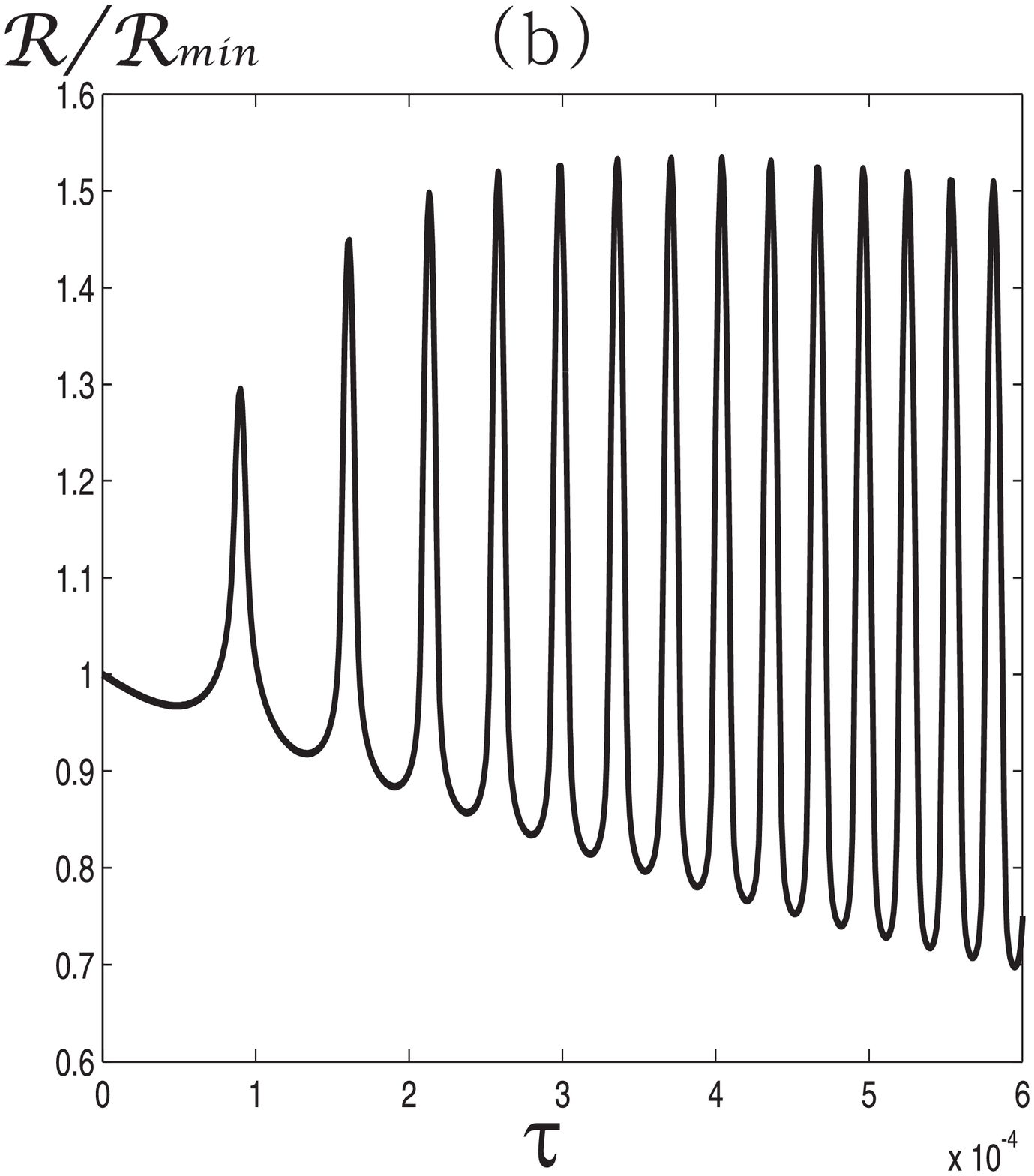}}
\end{center}
\end{minipage}\\[5mm]
\end{tabular}
\caption{$R/R_{min}$ as functions of $\tau$ in the modified Type-II $f(R)$ models for
$\tau_{ch}=10^{-3}$
 and $(g_{II},\beta)$=
$(1,25)$ and $(10^{-2},20.4)$ for (a) and (b), respectively.
}
\label{fg:6}
\end{figure}
\end{center}

In Fig.~\ref{fg:5}, we display $R/R_{min}$ as functions of $\tau$ in the modified Type-I $f(R)$ models with $n=2$, where
$R_{min}=\kappa^{2} T_{0}(1+\tau/\tau_{ch})$.
The large values of  $g_I$  and $\tau_{ch}$ correspond to the smaller amplitude of $R/R_{min}$, and vise versa.
If $g_I$ is small enough (see Figs.~\ref{fg:5}b and \ref{fg:5}d), the divergent behavior of the original viable $f(R)$ part is more efficiency,
and the amplitude increases
 until the effect of the $R^{2}$ part is big enough.
 In Fig.~\ref{fg:6},  we illustrate the similar results as Fig.~\ref{fg:5} for the modified Type-II $f(R)$ models with
 two different values of $g_{II}$, related to the background density. From the figures, we find that
the amplitude in Fig.~\ref{fg:6}b increases as that in Fig.~\ref{fg:5}a, and it is almost a constant in Fig.~\ref{fg:6}a.
It is clear that the amplitude in the Type-II  models is more sensitive than the
Type-I ones when we change $g_I$.

In the physical system, we can estimate the order of $g_{II}$ in different background densities
in the modified Type-II $f(R)$  models. Under the natural selection of the mass factor $M\simeq M_{pl}$, leading to
\begin{eqnarray}
\label{eq:ExpEstimate}
\beta =\kappa^{2} T_{0}/R_{ch} \gtrsim ln\left( \frac{\lambda M_{pl}^{2}}{R_{II}} \right) \sim 280\,,
\end{eqnarray}
we find
that $g = O(1)$, where we have assumed $\lambda \sim O(1)$ and $R_{ch}/M_{pl}^{2} \sim \Lambda /M_{pl}^{2} \sim 10^{-121}$. This is a very low energy density since the physical density, such as that in the inner galaxy regime, is much larger.
As a result, it is hard to detect the oscillation behavior and scalar mode of gravitational waves in the Type-II $f(R)$  models.
On the other hand, the curvature oscillation and scalar mode of gravitational waves
could still be detected
in the modified Type-I $f(R)$ models.  Note that the amplitude  depends on both $g_I$ and  $\tau_{ch}$ (see Figs.~\ref{fg:5}a and \ref{fg:5}c). The galaxy collision takes billions years ($t_{ch} \sim 10^{16} sec $ and $\tau_{ch} \sim 10^{14} \gg 1 $ for $n=2$ and $\beta = 10^5$), so that $\delta R / R_{min} \rightarrow 0$ when $g_{I} \gtrsim O(1)$.
The curvature oscillation and scalar mode of gravitational waves can only be observed for a small value of  $n$ as it corresponds to $g_{I} \lesssim O(1)$,
which can be  estimated  by using Eq.~(\ref{eq:PrevHSSPara}) with $R_{ch} \sim \Lambda \sim 10^{-29} g cm^{-3}$.
For example, for the inner galaxy  (sun) with the density $\rho_{m} \simeq 10^{-24} g/cm^{3} (\rho_{\odot} \simeq 1.4 g/cm^{3})$,
one finds that  $n \leq 11 (= 1)$.

In the above discussions, we have assumed  the flat space-time. Now, we would check
  whether this approach is valid or not when the curvature becomes large.
  If the high density region is described by the locally homogeneous and isotropic FRW metric,
  one has the form of $g_{\mu \nu} = \mathrm{diag}(-1,a^{2},a^{2},a^{2})$, where $a$ is the scale factor. Then, the gravitational field (Friedmann) equation is given by
\begin{eqnarray}
&&3\left(1+ f_{R} \right) H^{2}=\kappa^{2} \rho_{m} + \frac{1}{2} \left( Rf_{R} - f \right) - 3H \dot{f}_{R},
\label{eq:Friedmann1}
\\
&&-2 \left(1+ f_{R} \right) \dot{H} = \kappa^{2} \left( \rho_{m} - P_{m} \right) + \ddot{f}_{R} - H \dot{f}_{R},
\label{eq:Friedmann2}
\end{eqnarray}
with $R=6\left( \dot{H} +2H^{2} \right)$, where $H\equiv \dot{a}/a$ is the Hubble constant.
In Eq. (\ref{eq:Friedmann1}), the second term of  is approximately equal to a constant curvature of $\lambda R_{I}$, which is much smaller than the first term ($\kappa^{2} \rho_{m}$), while the third term is
\begin{eqnarray}
-3H\dot{f}_{R}=-3H \gamma^{-1}f^{\prime}_{R} \sim -H \sqrt{R_{I}} \left( \frac{R_{I}}{\kappa^{2} T_{0}} \right)^{n+1} \left( \frac{\kappa^{2} T_{0}}{R} \right)^{2n+2} \left( \frac{R}{\kappa^{2} T_{0}} \right)^{\prime}.
\label{eq:EstimH2}
\end{eqnarray}
As the large curvature oscillation corresponds to a small $g_{I}$, we can estimate each quantity in Eq. (\ref{eq:EstimH2}).
Since Eq. (\ref{eq:PrevPLNum})  approximately describes a simple harmonic oscillator $g_{I} z^{\prime \prime} +z -\left( 1+ \tau / \tau_{ch} \right) = 0$
for $g_{I} z^{2n+2} \gtrsim 1$, where $z=R/\kappa^{2} T_{0}$, leading to the oscillating frequency $\omega = g_{I}^{-1/2}$.
Without loss of generality, we take $g_{I} z^{2n+2} \sim 1$. Consequently,
the order of the oscillation amplitude of $H^2$ in Eq. (\ref{eq:EstimH2}) is $-3H \dot{f}_{R} \sim - (H/M_{pl}) R(\kappa^{2}T_{0}) R_{I} \ll R_{I} \ll \kappa^{2} \rho_{m}$.
Similarly, in Eq. (\ref{eq:Friedmann2}), the oscillation amplitude of $\dot{H}$ is dominated by $\ddot{f}_{R}$, which has the same order as the  curvature oscillation amplitude:
\begin{eqnarray}
\frac{\dot{H}}{\kappa^{2} T_{0}} \sim \frac{\ddot{f}_{R}}{\kappa^{2} T_{0}} = \frac{ \gamma^{-2}}{\kappa^{2} T_{0}} f^{\prime \prime}_{R} \sim \left( \frac{\kappa^{2} T_{0}}{R} \right)^{2n+2} \left( \frac{R^{\prime \prime}}{\kappa^{2} T_{0}} \right) \sim \frac{R}{\kappa^{2} T_{0}}.
\label{eq:EstimdotH}
\end{eqnarray}
As a result, the singularity behavior comes from the small amplitude and high frequency of the scale factor oscillation.
Hence, the result still holds under in the flat space-time limit. Explicitly, the covariant derivative yields  the same result as the partial derivative,
$e.g.$, $\square R = -\left( \ddot{R} + 3H \dot{R} \right) \simeq -\ddot{R}$.

\begin{center}
\begin{figure}[tbp]
\begin{tabular}{ll}
\begin{minipage}{80mm}
\begin{center}
\unitlength=1mm
\resizebox{!}{6.5cm}{\includegraphics{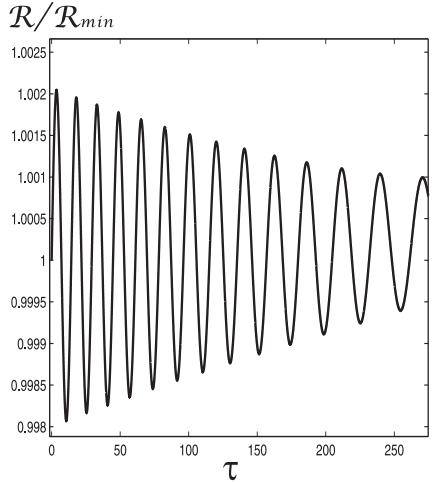}}
\end{center}
\end{minipage}
\end{tabular}
\caption{ $R/R_{min}$ Vs. $\tau$
when $\tau_{ch}=-1100$ is a negative number with $n=2$ in the original
Type-I $f(R)$ models without the $R^{2}$ term.
}
\label{fg:7}
\end{figure}
\end{center}

Finally, it is worth to mention that the amplitude of $R/R_{min}$ would be converge if $\tau_{ch}<0$ in the original
Type-I $f(R)$ models without the $R^{2}$ term as shown in Fig.~\ref{fg:7}. This result can be seen as the time reversion of $\tau_{ch} >0$. This kind of singularity does not exist in the cosmological evolution.

\section{Conclusions\label{sec:conclusions}}
We have demonstrated  that the singularity  appears in the finite time in the viable $f(R)$ gravity models. However, this singularity could be avoided by adding the $R^{n}/M^{2\left( n-1 \right)}$ term with $ 1 < n \leq 2 $ into the Lagrangian. We have explicitly shown
 the cases with the most reasonable parameters of $n=2$ and $M=M_{pl}$. Even though we can prevent the divergence, the oscillation behavior still exists.
  If the oscillating amplitude is large enough, it would be detected by cosmological observations. The oscillating effect
   in the different scale depends on the types of $f(R)$ models. In particular, in the Type-I models the effect can be induced in inner galaxy regime, but it is hard to distinguish the Type-II models from GR.

As the simplest example, the scalar mode of gravitational waves is a typical  phenomenon, which is different from GR, and it has the same origin as the curvature oscillation.
As shown, the Type-I viable $f(R)$ models can give  good sources of the scalar gravitational waves when $n \leq 11$. However, if $n$ is too large
(the case of $n>12$  leads to  $g \sim 1$ with the inner galaxy regime) or the background is too dense (denser than the solar density), the amplitude stays in a stable small fluctuation regime and it cannot be the source of the scalar mode. This behavior does not appear in the Type-II viable $f(R)$ models, because the factor $g_{II}$ is too large when $\kappa^{2} T_{0}> 280 R_{ch}$. This is still  a very low density regime and hence, the energy of the scalar mode of gravitational waves~\cite{arXiv:1106.5582} cannot be emitted.

\begin{acknowledgments}
The work was supported in part by National Center of Theoretical Science
and  National Science Council (NSC-98-2112-M-007-008-MY3) of R.O.C.
\end{acknowledgments}

\end{document}